\documentstyle[12pt,aasms4,flushrt]{article}

\slugcomment{submitted to The Astrophysical Journal}

\lefthead{A. Gonz\'{a}lez}
\righthead{Tidal shear}

\begin{document}

\title{Large-scale tidal fields on primordial density peaks ?. II\\
       Alignment of cosmic structures. }

\author{Alejandro Gonz\'{a}lez}
\affil{Coord. de Astrof\'{i}sica, Instituto Nacional de 
       Astrof\'{i}sica, Optica y Electr\'{o}nica. \\
       A.P. 51 y 216, C.P. 72000, Tonantzintla, Puebla, M\'{e}xico.}




\begin{abstract}
We show that the primordial density field imposes certain
degree of coherence in the orientation of density perturbations. 
We quantify the scale of coherence and show that is significant 
on scales of at least $30-40h^{-1}$Mpc, being more important in 
density fields with flat spectrum. Evidence is also presented that 
the reason for this coherence is that the long density waves are 
the dominant part of the superposition of waves which build the 
large-scale density peaks. As a consequence of this, small-scale 
peaks tend to follow the configuration of their host large-scale 
perturbations. Different types of alignments are investigated. It 
is shown that alignment of the major axes of neighbouring peaks is 
more prominent around the highest peaks than around the lower ones. 
Alignments between the major axes of peaks and the radius-vector 
joining their centre with the centre of a high peak were not observed. 
Evidence is presented that peaks develop tails extending to 
neighbouring peaks, as predicted by Bond (1987a, 1987b) and 
Bahcall (1987).

\end{abstract}


 

%

\section{Introduction}

Several observational works have reported different kinds and degrees of 
non-random orientation of cosmic structures. For example, on large-scale 
Gregory et al. (1981) showed that the major axis of the Perseus-Pisces 
supercluster coincides with the peak of the distribution of the position 
angle of galaxies. Djorgovski (1983) found the same effect for the Coma
cluster and similar results have been reported for the Local supercluster
(MacGillivray and Dood, 1985; Flin and Godlowski 1986; Kashikawa and Okamura,
1992). There also exist evidence for cluster-cluster major axes alignment over 
tens of megaparsecs (e.g. Binggeli, 1982; West, 1989; Plionis, 1994). On 
smaller-scales, a possible morphology-orientation effect has been reported 
by Lambas, Groth and Peebles (1989) and Muriel and Lambas (1992), who proposed
that the alignment of elliptical galaxies reflects the primordial
orientation of density maxima. Other point of interest concerns the evidence 
that first ranked galaxies, e.g cD galaxies, in linear clusters show the 
same orientation as their parent structures (Sastry, 1968; Carter and 
Metcalfe, 1980; Trevese, Cirimele and Flin, 1992).

On the theretical side there are numerical simulations which suggest that
that alignment of structures arise in different cosmological models wherever 
filaments are formed (e.g. West, 1989; West, Dekel and Oemler, 1989; Dekel, 
1989; West, Villumsen and Dekel, 1991; Plionis, Valdarnini and Jing, 1992).
Moreover, there also exist increasing evidence that some coherent 
patterns in the orientation of density perturbations do exist in the 
primordial density field (Bond, 1987a, 1987b; Barnes and Efstathiou, 1987; 
Gonz\'{a}lez, 1994, 1997; Bond, Kofman and Pogosyan, 1996), discounting the 
action of small-scale tidal fields of neighbouring galaxies. The alignment of 
galaxy clusters detected up to $30-50$Mpc would be difficult to explain via 
small-scale tidal fields. Instead, Bond, Kofman and Pogosyan (1996)and 
Gonz\'{a}lez (1997, AGI97) have shown that a high degree of coherence 
can be produced in the density field as the result of a constructive 
interference of long density waves which make rare peak-patches. This 
constructive interference would lead to alignment of density peaks.
The degree of coherence was found to be more prominent in density
fields characterized by a flat spectrum. The influence of the large-scale 
density field on small-scale peaks has been proposed as a possible 
generator of alignment of density peaks, and this in turn, to filamentary 
structures rehanced by the non-linear evolution of the field. 

The correlation of the density field on two different filtering scales, 
$R_{a}$ and $R_{b}$, assessed in terms of their rms ($\sigma_{a}$ and 
$\sigma_{b}$) was already exhibited by Padmanahban (1993) and West (1994) 
via the parameter
\begin{equation}
   \gamma_{ab}=\sigma_{ab}^{2}/(\sigma_{a}\sigma_{b}).
\end{equation}
For a field characterized by a power law spectrum $P(k)=Ak^{n}$,
\begin{equation}
    \sigma_{ab}^{2}\equiv < \delta_{a}\delta_{b}> = 
    \int P(k)W_{G}(kR_{a})W_{G}(kR_{b})d^{3}k,
\end{equation}
where $W_{G}$ is the Gaussian smoothing window function. Thus, the correlation 
parameter
\begin{equation}
\gamma_{ab}(n=1,0,-1,-2)=[\frac{2R_{a}R_{b}}{R_{a}^{2} + 
R_{b}^{2}}]^{s},
\end{equation}
with $s=2,3/2,1,1/2$ respectively, is $\gamma_{ab}=1$ only when the same 
scale of filtering is used. On the contrary, 
$\gamma_{ab}\rightarrow 0$ indicates that the correlation decays. A  
correlation is observed (Figure 2 in West, 1994) between small and 
large scales, being more 
important ($\gamma_{ab}>0.5$) for flat spectra density fields, up to 
approximately $R_{f}=5-12$. For steep spectra, the correlation is 
only important in the nearest neighborhood of the peaks up to $R_{f} 
< 3$. These results are similar to those found in AGI97, where the strenght 
of the tidal field was evaluated through the Frobenius norm. For flat spectra, 
long waves as well as short ones help to build the density peaks, but long 
waves should be the dominant part. Long density waves create a coherent 
effect which can persist out to large distances. If one filter the density 
field at a larger scale, the coherent effect dictated by the direction in 
which the large-scale density peak is oriented should be reflected in their 
internal regions. The influence of the large density waves extends not only 
to the lower-scale peak in question, but also to the neighbouring peaks. 
Thus a correlation in the orientations of small-scale peaks is also expected.
The goals of the present paper are to provide evidence that this latter
statement does happen, and to propose that different kinds of alignments
of cosmic structures could have a common primordial origin, as a result 
of the constructive interference of density modes of long wavelenght. In 
Sect. 2 orientations around the highest and low peaks are analized, as 
a function of the distance and the spectral index. We present three 
statistical test to prove alignment of density peaks. Radial aligment is 
also investigated. Sect. 3 focuses in the study of large-scale alignmemt 
of the type ``cD-cluster'', ``galaxies in clusters'' and ``tails'' 
connecting density peaks. In Sect. 4 we present a discussion of our 
results and the conclusions.     

\section{Alignments and its correlation scale}

In this section, we address the study of a possible alignment of peaks 
with their neighbours, for a density field smoothed with $R_{f}\equiv 1 
(\approx 0.66$Mpc). The orientations of all density peaks will be refered 
to the orientation of the highest peak in the density field. This choice 
is motivated because it suggests an important coherent superposition of 
density waves on and around the position of the highest peak. Thus, we 
assess the cosine of the angle subtended by the major axes of the inertia 
tensor of the highest peak and those of its neighbouring perturbations. 
In order to explore an alignment of peaks as a function of the distance,
we consider all those density peaks located within shells of equal thickness 
$D=10R_{f}\approx 7$Mpc up to a distance of $40R_{f}\approx 30$Mpc. Larger 
distances were not considered because of the periodicity of the box at 
$64R_{f}$. 

Three simple statistical tests are performed to detect and quantify any 
orientational anisotropy of peaks. {\it (a)} The mean value of $\cos\theta$ 
and its error in mean. Parallel and antiparallel alignments are not 
distinguished. For randomly oriented perturbations the distribution 
of $\cos\theta$ shall uniformly be distributed between $0$ and $1$ with 
mean value $0.5$. {\it (b)} The ratio between the number $N_{<0.5}$ of 
peaks with $\cos\theta < 0.5 $ and the number $N_{>0.5}$ with $\cos\theta 
> 0.5$. An excess of $N_{>0.5}$ over $N_{<0.5}$ would indicate the existence 
of alignments. {\it (c)} We also quantify the probability $P_{KS}$ according 
with the Kolmogorov-Smirnov test for the null hypothesis that the 
$\cos\theta$ distributions have been drawn from a uniform one.

Figure 1 shows for each of the spectral indexes, the distribution of peaks 
in the plane $R_{f}-\cos\theta$, and the histogram of the distribution in 
orientations of the peaks within the two first shells of radius $10R_{f}$.
These histograms and the statistical analysis summarized in Table 1, 
indicate the existence of alignment effects in the initial density field. 
While all the models exhibit non neglegible alignments within the first
shell to a distance up to $\approx 6-7$Mpc, for $n=1,0$ the alignment seems 
to persist very weakly beyond this distance. For flat spectra, the statistical 
significancy of the alignment extends up to at least $15-25$Mpc.

At the scale of clusters ($Rf\equiv1\approx 10$Mpc), for $n=-1,-2$, the 
alignment clearly extends up to at least $30$Mpc. At this point, it is 
interesting to remark that in the N-body simulations by West, Villumsen 
and Dekel (1991) on the issue of cluster-cluster alignments, and Matarrese
et al. (1991) 
who used groups of particles in a CDM model with non-Gaussian initial 
conditions, alignment effects were found up to a scale of $40-60h^{-1}$Mpc. 

Before discussing the possible origin of the alignments and other 
consequences, we comment on three additional test which were performed 
in order to confirm the alignment of density perturbations.

\subsection{Testing the Alignment Effect}

We have identify the following three main sources of error;

\begin{enumerate}

\item The orientation of density peaks is defined in terms 
      of the main axes of the deformation tensor. A straightforward test 
      consisted in recalculating it with a higher resolution. Any 
      considerable change of any of the elements of the inertia tensor 
      would yield changes in the orientation of its main axes. Even when
      small changes in the position of the peaks in the plane 
      $N_{pk}-\cos\theta$ were observed, the overall distribution 
      presented no considerable changes. In any case, changes in the
      orientation would not produce systematic alignments. 

\item The goal of the second test is to discard hide systematic errors. Two 
      realizations were performed for each of the spectral indexes. In the 
      first realization we identified the direction defined by the alignment
      of the major axes of two high peaks: the highest peak and other 
      placed to a distance $\sim 4R_{f}$. This direction approximately 
      defines the direction of alignment of the rest of the peaks, and 
      is refered relative to the cartesian coordinates of the $64^{3}$ grid.  
      We repeated this for a second realization. The same scale of alignment 
      was obtained in both realizations, however, the directions of alignment 
      were different.

\item The third test aims to probe the 'isotropy' of the effect, i.e. 
      the alignment has the same statistical significance when the 
      orientations are calculated respect to another high peak, instead
      of using the highest one. We test both distributions against randomness 
      in the sample using the Kolmogorov-Smirnov test for the null hypothesis 
      that both data sets are drawn from a uniform distribution. Note that, if 
      we choose by chance, another peak nearly aligned with the highest 
      one, then the distribution showed in Figure 1 should only suffer little 
      changes. Contrarily, if a peak is chosen which is not aligned, then the 
      distribution will be clearly modified. In that case, we would be proving
      that alignments ocurr around high peaks. Figure 2 and Table 2 display 
      the results for the case $n=-2$.
\end{enumerate}

The results of these tests strongly suggest that the alignments 
are not a product of numerical artifices. If this were so, the same degree 
of alignment, or even absence of it, would be produced in all the models 
independently of the spectral index. On the contrary, there is a clear 
dependence on the flatness of the spectrum: flat spectra show stronger 
alignment effects than the steeper ones. The scale of coherence for the 
former of these is also larger, extending up to $15-20R_{f}\approx 15$Mpc. 
Further, the fact that the degree of alignment depends on the distance to the peak 
of reference it would also be difficult to explain by numerical systematics.  
It should also be taken into account that the influence of the large-scale 
density field on the deformation tensor at the position of the small-scale 
peaks, quantified trough the Frobenius norm in AGI97, is larger 
for flat spectra, $\approx 10R_{f}$: a scale which is reasonable consistent 
with the scale of alignments we have just found. The main difference between 
a flat and a steep spectra lies in the the superposition of the type of 
density waves which form the peaks. In steep spectra, like $n=1,0$, the 
superposition is dominated by short density waves. Such a superposition only 
locally affect the density peak; presumably the wavelength of the density waves 
is smaller than the separation of density maxima. In flat spectra, the long 
density waves are the most important. If their wavelength is larger than the 
mean separation of peaks, then it would lead to a coherent coupling between 
density perturbations.

\subsection{Alignments as a function of the height of peaks}

One way of exhibiting the influence of long density waves in the field and 
their correlation with the short waves, is through the comparison between the
distribution of orientations of neighbouring peaks surrounding high 
peaks, with that for low ones. We restrict this analysis for 
$n=-2$, where a possible difference should be easily appreciated.

In Figure 3 we have plotted the distribution of orientations around the highest 
peak ($\nu > 3\sigma$); also shown is the distribution of orientations for low
peaks ($\nu < 2\sigma$) refered to the highest one of the sample. By comparing 
these two distributions, we observe that low peaks surrounding low peaks, poorly 
follow the distribution of the highest one. Low peaks are approximately uniform 
distributed. Figure 3 also shows the distribution of orientations of 
neighbouring peaks for some of the highest peaks ($\nu > 3.8$)
and around some low peaks ($\nu < 1.5\sigma$). All the neighbouring peaks
within a distance of $8R_{f}$ from the peak of reference are
considered. By comparing these figures, one first see that the influence 
of the highest peaks over the orientation of their neighbours is more 
important than the influence produced by the lower peaks. This is 
observed from the more marked tendency of the neighbours to cluster near 
$\cos\theta$. This fact, therefore, reflects the importance of long density 
waves. Second, the excess of short waves in $n=1$ leads to no correlation. 

We have analysed in further detail the individual characteristics of the 
neighbouring peaks, for each of the cases of Figure 3. The analysis 
included the study of the spatial distribution of the neighbouring peaks, by 
means of which an excess of peaks along the main axes of a high peak  would be
detected. The space around each peak was divided as in Figure 4. The height 
of the peaks and the distance of the neighbouring peaks were considered. 
The results can be summarized as follows; for $n=-2$ where the highest peak 
was $\nu\approx 4$ we analysed 30 cases. The average number of neighbours 
was 13.1. The number of peaks located within the two $120^{o}$ cones, at both 
sides of the peak, was in average 6.3, whereas within the two $60^{o}$ cones
was 6.8, consistent with an isotropic distribution. An important difference 
is noted when one considers the number of peaks according withwhether they are 
higher or lower than the reference peak. Seven out of our 30 cases had two or 
three peaks of height approximately equal or slightly lower than the peak of 
reference, within the $120^{o}$ cones. 

For $n=1$, 64 high peaks were analysed. In this sample, the highest
peak was $\nu\approx 2.8\sigma$ and the average number of neighbours was 
16.8. Their distribution in space is also consistent with isotropy
as it is reflected by the values of the average number of peaks within the 
region limited by the cones, 8.2, and  8.6. 

\subsection{Radial Alignments}

In order to detect any radial alignment around the high density peaks we assess 
the cosine of the angle subtended by the radial vector, which joins the 
center of the highest peak with the center of its neighbours, and the major 
axis of the inertia tensor. Only the cases $n=1$ and $n=-2$ were considered.
No signals of radial alignments were detected as is suggested by the  
distribution of peaks in the examples of Figure 5. 

The absence of radial alignments is probably not surprising, because the 
superposition of density waves is not isotropic as inferred from the tendency 
of the peaks toward triaxiality. The coherent wave bunching which forms the 
density peaks, make them largest along the long axis and smallest along the 
short axis. As the long density waves dominate the field, according to the 
parallel alignment earlier detected, they impose a preferential direction 
along which lower peaks will tend to align so that they point each other, 
specially if they are not very far away. 

\section{Large-scale Fossil Alignments}

There is a possible immediate consequence of the parallel alignments observed 
in flat spectra: once the density field is smoothed at a larger scale is
conceivable that the new large-scale peaks should preserve certain information
about the 'average' triaxiality of the small-scale peaks from which they 
arise, especially if the two scales are not exaggerately different. Consider, 
for example, that the density field instead of being smoothed on a scale $R_{f}$ 
is smoothed on a scale $R_{f} + \delta R_{f}$. Intuitively one expect that the 
characteristic of the peaks such as positions, shapes and orientations are 
similar in both situations. In other words, the orientation of small-scale 
density peaks should trace the orientation of the large-scale inhomogeneities.
Following this idea we analyse two possible manifestations of alignments.

\subsection{ A cD-cluster Alignment effect ?}

A cD galaxy approximately corresponds to a Gaussian smoothing of $R_{f}\sim 
2-3h^{-1}$Mpc. This filtering scale is not much smaller than the 
$8-10h^{-1}$Mpc appropriate for clusters, where smaller means that those
waves of length $\sim 8h^{-1}$Mpc make a significant contribution to the 
rms density field on the scale of cDs. Therefore, a correlation in the 
orientation of the density perturbations on these scales would suggest a 
primordial origin for the tendency of cD galaxies to be aligned with 
their host cluster (Carter and Metcalfe 1981). We now address this
possibility for a density field with $n=-2$.

We calculate, in a similar way to that in AGI97, the relative orientation 
between the density peaks of the field smoothed on scales of $2h^{-1}$Mpc 
and $8h^{-1}$Mpc. The major axes of the deformation tensors are once again 
used to ascertain the position angles. The positions of the small-scale peaks 
relative to the center of the large-scale fluctuation are also required. We 
further assume that the small-scale peaks belong to the ``cluster-scale'' 
fluctuation if they are contained within a shell of radiu $8R_{f}$.

Figure 6 shows one successful examples of the distributions of the 
orientation of the major-axes of small-scale peaks relative to the major 
axis of the large-scale host fluctuation. A tendency of the small-scale 
peaks to be aligned with the larger structure was found only in 6 out of 
32 cases analysed. Though this result confirms a non negligible deviation 
from random orientations, we could not obtain a small-scale peak placed in 
the center of the cluster-scale peaks so as to mimic a cD galaxy. The position 
of the small-scale peaks in general presented considerable displacements from 
the center of the cluster-scale peaks. In any case, it is noteworthy that 
the orientation of the major axes of the small-scale density perturbations 
follow the orientation of the large-scale peaks, similar to those involving 
the brightest elliptical galaxies in clusters (e.g. Sastry, 1968; Carter 
and Metcalfe, 1980; Rhee and Katgert, 1987).

\subsection{Asymmetric tails of density peaks ?}

We now briefly explore the proposal first made by Bond (1987a, 1987b) and 
Bahcall (1987) concerning the existence of long tails around high density peaks, 
which could connect them with their near neighbours. Although the detection of 
tails along the major axes of the peaks is hard to perform for the whole field, 
mainly due to the different threshold values of the density contrast to explore,
we provide some supporting evidence of their existence.

In the detection of peaks tails we proceed by interpolating the density at
points surrounding the peaks and then plotting isodensity surfaces around them,
looking at their shape and length along their major axes. Isodensity 
contours around neighbouring peaks will be trivially connected if a low 
density threshold is chosen. We are interested in detecting 'bridges' between 
peaks when a high density threshold ($\nu < 15\%\nu_{peak}$) is considered. 
This means, isodensity contours extended along the main axes of high peaks.

Figure 7 shows an example of the positive results in searching for tails, for
$n=-2$. 20 peaks were analysed in this way but only those (3 cases) peaks 
with a near (separated a distance $< 10R_{f}$) neighbour of similar height 
presented tails. This result further support the idea that density modes produce
a coherent orientation along the major axes of high peaks when their
wavelengths are larger than the mean separation of peaks, which in the
case $n=-2$ is $5R_{f}$ for peaks of arbitrary height, and $\approx 8R_{f}$ for
peaks with $\nu > 2.5$ (Gonz\'{a}lez, 1994).  We can argue for these peaks that, 
as the density field evolves, the small-scale peaks are carried along with the 
large-scale distribution of mass, which should reinforce their initial 
parallel alignment.

\section{Discussion and Conclusions}

We have found evidence of parallel alignment of density perturbations
on different scales. We found for example that the small-scale density 
peaks follow a similar orientation to that of the cluster-scale major axis. This
effect is more important for flat spectra. When 
filtering on cluster-scales is considered, the alignments extend up to 
$\approx 30h^{-1}$Mpc, in agreement with the observational evidence of 
cluster-cluster alignments (.e.g. Plionis 1994). Further, the scale of 
correlation deduced from the analysis of changes in the deformation tensor, in
AGI97, and the scale of alignments reasonably agree with the 
theoretical scale of correlations. Similar agreement is found with the 
strength of the tidal influence on the peaks, estimated through the Frobenius 
norm.

Evidence was given that flat spectra constitute a coherent density field:
i.e. the orientation of small-scale peaks follow the orientation of the 
large-scale fluctuations. Whereas the correlations of the density field 
smoothed at two different scales are a natural consequence of Gaussian density 
fields, the alignments are related to the flatness of the spectrum. This is 
clearly indicated by the absence of alignments on large scales for steep spectra.

Based on these results we conclude that an important degree of large-scale 
alignment of the cosmic structure is of primordial origin, in agreement with 
Bond, Kofman and Pogosyan (1996). Some numerical simulations 
(e.g. Melott and Shandarin 1989; Kofman et al. 1992) have shown that
the development of filamentary structures is a generic feature of density fields
with flat spectra. Our results, which have proved to be quite 
sensitive to the type of spectrum of the density field, indicate therefore than
alignments and filamentary structures are two correlated aspects. Bond (1987a,
1987b) and Bahcall (1987)have visualized the filamentary structure of the 
Universe as primordial density peaks with long tails connecting peaks. These 
bridges would be dynamically enhanced by the non-linear evolution of the field. 
This idea can account not only for the filamentary structure but also for the 
alignment of structure. We have provided evidence of the existence 
of such long tails in flat spectra and coherence in the orientations of peaks. 
As another possible consequence of these tails, which deserve more attention, 
is the anisotropic shape of the cluster-galaxy correlation function 
(Bahcall 1987). Both the spatial cluster-cluster correlation function and the 
cluster-cluster alignment joined by bridges can give a natural explanation for 
such anisotropy.

There are two important factors which will determine the final orientation of
collapsed objects which form at the sites of density peaks; a) the tidal 
stress, which is produced by the surrounding distribution of mass, b) and the 
initial orientation of the density perturbation relative to the tidal field. 
The main effect of the tidal shear is to change 
the orientation of the deformation tensor, which leads to significantly modify 
their initial orientation and shape (e.g. van de Weygaert and Babul, 1994).
Tidal stress is stronger for the cases $n=-1,-2$.

As the density field enters non-liner regime, different Fourier modes get
couple,  giving an extra dynamical contribution to their initial statistical 
correlation both spatial and orientational. The initial orientational 
anisotropy is further rehanced by the intrinsic triaxiality of peaks which 
follow the Zel'dovich predictions. If the peaks possess long tails connect 
them with near neighbours, then the simultaneous collapse with the 
large-scale density perturbations would reinforce the anisotropy of the 
initial chained peaks. The signals of alignments found for flat spectra
reflect the important contribution of large-scale fluctuations, which
both induce strong tidal fields and make the collapse of cluster
to follow each other in quicker succession. From this point on, 
it is necessary to perform numerical simulations in order to elucidate whether 
the non-linear coupling of density waves is not enough so as to erase some
of their initial coherent superposition.

\clearpage

\figcaption[fig1.ps]{Distribution of peaks respect to the highest 
           peak in the sample, as a function of their relative orientation 
           and their separation. The histogram shows the 
           distribution of peaks with separations $< 15R_{f}$. A significant 
           alignment effect is observed up to $\approx 20R_{f}$ for 
           $n=-1,-2$.}

\figcaption[fig2.ps]{A comparison of the orientation distribution 
           of neighbouring peaks, calculated from any of the highest 
           peaks.}

\figcaption[fig3.ps]{{\it (a)} The distribution of orientations 
           of high peaks $\nu >3.8\sigma$ --relative to the highest 
           one-- in a model with $n=-2$. {\it (b)} The 
           distribution of low peaks $\nu < 1.5\sigma$. Also shown 
           are the orientations of neighbouring peaks for some of 
           the highest and low peaks. }

\figcaption[fig4.ps]{The space around the highest  peak is divided in 
           two areas as illustrated here.}

\figcaption[fig5.ps]{No radial alignments are detected; the orientation 
           of the major axes of peaks relative to the radius vector 
           joining their centre with the centre of the nearest high peaks.}

\figcaption[fig6.ps]{{\it cD-cluster} alignments.}

\figcaption[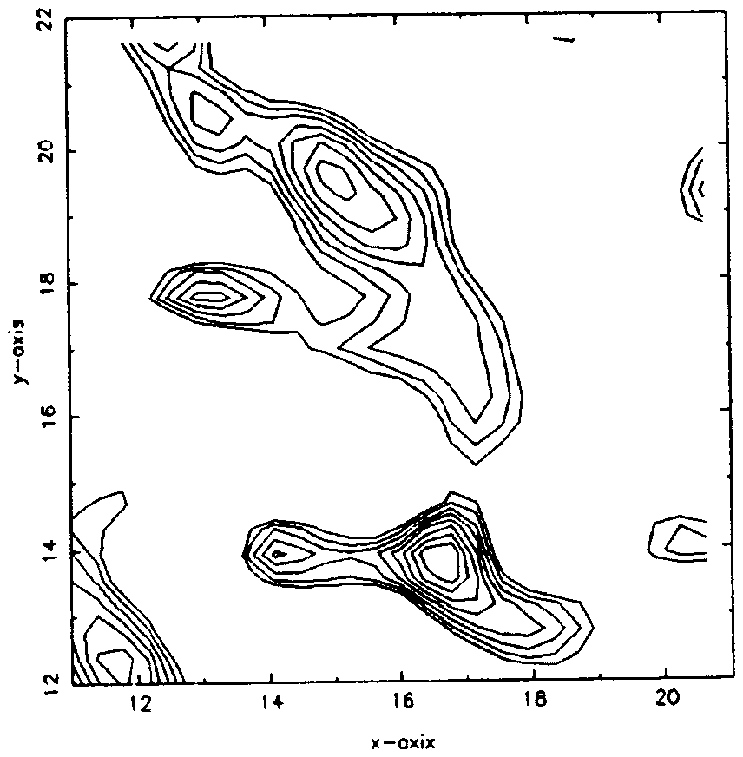]{One case in which a long tail connect high 
           density peaks.}

\clearpage 

\begin{deluxetable}{lrrr}
\tablewidth{33pc}
\tablecaption{Statistics for Alignments}
\tablehead{
\colhead{ Spectral index n }      & \colhead{ $<\cos\theta>$ }  &
\colhead{ $N_{<0.5}/N_{>0.5}$ }   & \colhead{ $P_{KS}$       }  \\
\colhead{ $D (h^{-1})$Mpc }       & \colhead{                }  &
\colhead{                 }       & \colhead{                }  }

\startdata
$n= 1$              &                   &              &               \nl
$0\leq D \leq 10$   & $0.572\pm0.0219$  & $198/288$    & $< 10^{-16}$  \nl
$10\leq D \leq 20$  & $0.539\pm0.0012$  & $1052/1212$  & $0.05$        \nl
$20\leq D \leq 30$  & $0.492\pm0.0312$  & $1274/1348$  & $0.27$        \nl
$30\leq D \leq 40$  & $0.485\pm0.0241$  & $316/323$    & $0.46$        \nl
$n= 0$              &                   &              &               \nl
$0\leq D \leq 10$   & $0.615\pm0.0219$  & $143/278$    & $< 10^{-16}$  \nl
$10\leq D \leq 20$  & $0.579\pm0.0012$  & $952/1132$   & $< 10^{-3}$   \nl
$20\leq D \leq 30$  & $0.522\pm0.0312$  & $1035/1114$  & $0.13$        \nl
$30\leq D \leq 40$  & $0.505\pm0.0241$  & $153/168$    & $0.23$        \nl
$n=-1$              &                   &              &               \nl
$0\leq D \leq 10$   & $0.684\pm0.0310$  & $94/196$     & $< 10^{-17}$  \nl
$10\leq D \leq 20$  & $0.651\pm0.0112$  & $486/812$    & $< 10^{-8}$   \nl
$20\leq D \leq 30$  & $0.626\pm0.0731$  & $919/1158$   & $< 10^{-9}$   \nl
$30\leq D \leq 40$  & $0.582\pm0.0324$  & $385/469$    & $0.02$        \nl
$n=-2$              &                   &              &               \nl
$0\leq D \leq 10$   & $0.737\pm0.0119$  & $63/214 $    & $< 10^{-19}$  \nl
$10\leq D \leq 20$  & $0.715\pm0.012$   & $262/750$    & $< 10^{-9}$   \nl
$20\leq D \leq 30$  & $0.709\pm0.003$   & $422/1167$   & $< 10^{-5}$   \nl
$30\leq D \leq 40$  & $0.702\pm0.0241$  & $114/325$    & $< 10^{-7}$   \nl
\enddata
\tablecomments{Different statistical test to quantify the alignment 
               effects around the highest density peaks for power 
               law spectra.}
\end{deluxetable}

\begin{deluxetable}{lrrr}
\tablewidth{33pc}
\tablecaption{Alignments}
\tablehead{
\colhead{ Spectral index n }      & \colhead{ $<\cos\theta>$ }  &
\colhead{ $N_{<0.5}/N_{>0.5}$ }   & \colhead{ $P_{KS}$       }  \\
\colhead{ $D (h^{-1})$Mpc }       & \colhead{                }  &
\colhead{                 }       & \colhead{                }  }

\startdata
$n= -2$            &                   &             &         \nl
$0\leq D \leq 10$  & $0.724\pm0.0149$  & $80/194$    & $0.22$  \nl
                   & $0.717\pm0.0169$  & $68/206$    &         \nl
$10\leq D \leq 20$ & $0.705\pm0.0253$  & $293/757$   & $0.31$  \nl
                   & $0.715\pm0.0162$  & $318/732$   &         \nl
$20\leq D \leq 30$ & $0.695\pm0.0342$  & $493/1096$  & $0.37$  \nl
                   & $0.702\pm0.0142$  & $467/1122$  &         \nl
$30\leq D \leq 40$ & $0.687\pm0.0141$  & $105/334$   & $0.28$  \nl
                   & $0.694\pm0.0235$  & $128/311$   &         \nl
\enddata
\tablecomments{Statistical test to show the equivalence of the 
               alignments calculated from different peaks, in a 
               model with $n=-2$.}
\end{deluxetable}

\end{document}